\documentclass[prl,twocolumn,superscriptaddress,showpacs]{revtex4}
\usepackage{graphicx,amsmath,amssymb,bm}

\def\be{\begin{equation}}
\def\ee{\end{equation}}
\def\ba{\begin{eqnarray}}
\def\ea{\end{eqnarray}}
\def\bas{\begin{eqnarray*}}
\def\eas{\end{eqnarray*}}

\begin{document}

\title{Solution of the center-of-mass problem in nuclear structure
calculations}

\author{G.~Hagen} 
\affiliation{Physics Division, Oak Ridge National
Laboratory, Oak Ridge, TN 37831, USA} 
\author{T.~Papenbrock}
\affiliation{Department of Physics and Astronomy, University of
Tennessee, Knoxville, TN 37996, USA} 
\affiliation{Physics Division,
Oak Ridge National Laboratory, Oak Ridge, TN 37831, USA}
\author{D.J.~Dean} 
\affiliation{Physics Division, Oak Ridge National
Laboratory, Oak Ridge, TN 37831, USA} 

\begin{abstract}
The coupled-cluster wave function factorizes to a very good
approximation into a product of an intrinsic wave function and a
Gaussian for the center-of-mass coordinate. The
width of the Gaussian is in general not identical to the oscillator
length of the underlying single-particle basis. The quality of the
separation can be verified by a simple procedure. 
\end{abstract}


\maketitle

The atomic nucleus is a self-bound quantum many-body problem. Its
Hamiltonian is invariant under translations and rotations. Thus, total
momentum  and total angular momentum are conserved
quantities, respectively. When one chooses a single-particle basis to
solve the nuclear many-body problem, one is confronted with a
dilemma. There is no single-particle basis of states that are
simultaneous eigenstates of the momentum operator and the
angular momentum operator. For calculations of finite nuclei
within the nuclear shell model, one usually chooses a spherical
single-particle basis. Such a basis obviously breaks translational
invariance, and the consequences have been addressed in numerous
papers (see,
e.g.~\cite{Elliott55,Lipkin58,Vincent73,Lawson74,McGrory75,Whitehead}).

Only two rigorous solutions to the problem are known. The first
approach deals with the problem in translationally invariant Jacobi
coordinates. We refer the reader to
Refs~\cite{Bishop,Nogga,Bacca} for recent applications. Due to
the factorial scaling of the required antisymmetrization, this method
is restricted to few-body problems with $A<7$ or so. The second
approach is suitable for many-body problems but severely limits the
choice of the single-particle basis to the eigenstates of the harmonic
oscillator.  For the $A$-body system, one considers an $A$-fermion
basis that consists of all single-particle product states with
excitations up to and including $N\hbar\omega$. The corresponding
model space is called a full $N\hbar\omega$ space. In this space,
eigenstates of the translationally invariant nuclear Hamiltonian are
also eigenstates of the center-of-mass Hamiltonian $H_{\rm cm}
(\omega)$ (see Eq.(~\ref{Hcm}) below for a definition).  Note that the
Fock-space basis employed in this approach is not translationally
invariant, i.e. the total momentum is not a conserved
quantity. Rather, the complete $N\hbar\omega$ space ensures that all
eigenstates are products of an intrinsic state $\psi_{\rm in}$ and a
center-of-mass state $\psi_{\rm cm}$. The intrinsic states are
invariant under translations, while the wave function of the
center-of-mass coordinate is obviously not an eigenstate of the total
momentum. Thus, the factorization \be
\label{factor}
\psi=\psi_{\rm cm}\psi_{\rm in} 
\ee 
is central to the $N\hbar\omega$ space, and it is essential for a correct
treatment of the center-of-mass problem. Note that the particular form of the
center-of-mass wave function $\psi_{\rm cm}$ is
irrelevant~\cite{Vincent73,Whitehead}. Recent examples for this approach
are, e.g., the no-core shell-model calculations~\cite{Nav00,Nav09} and
the $0\hbar\omega$ shell-model calculations~\cite{Caurier,Otsuka}.

Unfortunately, the two rigorous approaches to the center-of-mass
problem scale exponentially with the number of active nucleons.
Furthermore, many nuclei of interest are very weakly bound, and the
oscillator basis does not provide the correct radial asymptotics.
Alternative wave-function-based approaches such as the coupled-cluster
method~\cite{Coe58,Coe60,Ciz66,Ciz69,Kuem78,Mi00,Dean04,Bar07}, the
unitary-model operator approach~\cite{Fujii04}, or stochastic shell-model 
approaches~\cite{SMMC,MCSM} scale more gently with increasing size of the
model space and the mass number. While these methods are also based on
a single-particle basis, they do not employ a complete $N\hbar\omega$
space. Thus, there is no analytical guarantee that the wave functions
computed by these methods also exhibit the
factorization~(\ref{factor}). For this reason, the results obtained by
such methods are occasionally viewed with skepticism. The concern
seems not primarily with the possible breaking of translational
invariance, but rather with the problem to quantify the errors that
might be involved. Recall that nuclear lattice calculations also break
translational invariance, albeit in a controlled way~\cite{Dean}.

It is the purpose of this Letter to eliminate these
concerns. Moreover, we present a simple tool (i.e., the computation of
expectation values of the generalized center-of-mass Hamiltonian) that
assesses to what degree the factorization~(\ref{factor}) is exhibited
in coupled-cluster calculations. In what follows, we consider the
$^{16}$O nucleus and demonstrate that the coupled-cluster wave
function is a product of an intrinsic wave function and a Gaussian for
the center-of-mass coordinate.

For the nucleus $^{16}$O we employ the low-momentum interaction
$V_{{\rm low} k}$~\cite{vlowk}, generated from Entem and Machleidt's chiral
nucleon-nucleon potential~\cite{N3LO} by imposing a smooth~\cite{smooth}
momentum cutoff $\lambda=1.8$~fm$^{-1}$. Our single-particle
basis consists of spherical oscillator states. The oscillator spacing
$\hbar\omega$ and the size of the model space (in terms of the number
$N$ of major oscillator shells) are the parameters of our
calculations. We employ the intrinsic Hamiltonian
\ba
\label{Hin}
H_{\rm in}&=&T-T_{\rm cm}+V \ , \nonumber\\
&=& \sum_{1\le i<j\le A} \left({(\vec{p}_i-\vec{p}_j)^2\over 2mA} +V(\vec{r}_i-\vec{r}_j) \right) \ .
\ea
Here, $T$ and $V$ denote the kinetic and potential energy operators,
respectively, and $T_{\rm cm}$ denotes the kinetic energy of the center
of mass.  The intrinsic Hamiltonian is manifestly invariant under
translations. We perform a spherical Hartree-Fock calculation to
obtain an optimized single-particle basis. This is followed by a
spherical coupled-cluster calculation~\cite{Hag08}. Clearly, the
Hartree-Fock single-particle basis in combination with the
coupled-cluster method in its singles and doubles approximation (CCSD)
does not employ a full $N\hbar\omega$ space. Thus, the 
separation~(\ref{factor}) is not
guaranteed from the outset of our calculations.

\begin{figure}[h]
\includegraphics[width=0.37\textwidth,clip=]{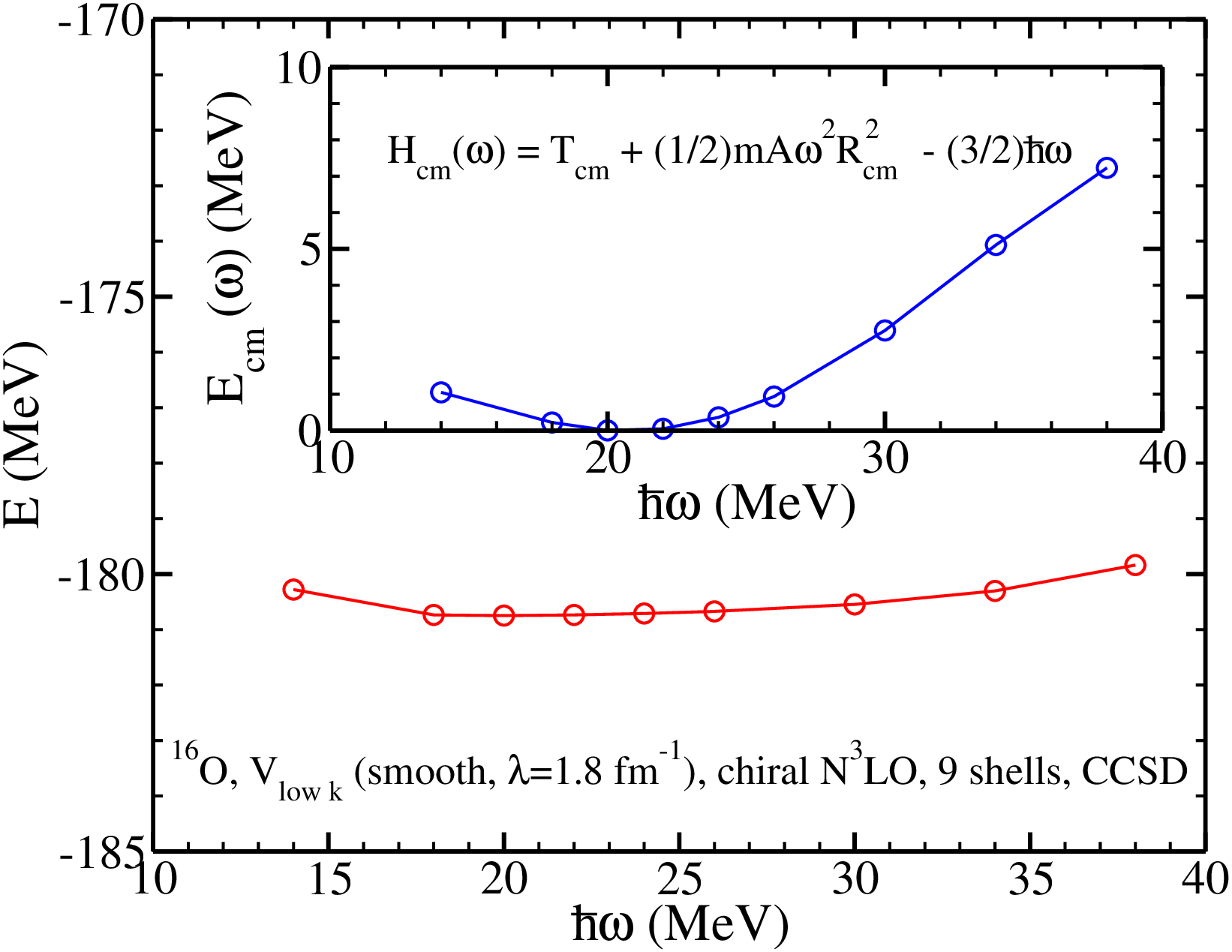}
\caption{(Color online) Ground-state energy (within CCSD) of
$^{16}$O with a low-momentum potential as a function of the oscillator
spacing $\hbar\omega$. The model space consists of nine major
oscillator shells. Inset: Expectation value $E_{\rm cm}(\omega)$ of the center-of
mass Hamiltonian with the standard frequency dependence.}
\label{fig1}
\end{figure}

Figure~\ref{fig1} shows the ground-state energy as a function of the
oscillator spacing for a model space consisting of $N=9$ oscillator
shells. The energy varies by about 1~MeV while the oscillator spacing
$\hbar\omega$ varies by more than a factor of two. This shows that the
ground-state energy is very well converged with respect to the size of
the model space. Note that we could easily employ larger model spaces
and improve the convergence of our results. However, this is not
necessary for the purpose of this Letter, and we refer the reader to 
established benchmark calculations~\cite{CCbench}.

Let us consider the generalized center-of-mass Hamiltonian
\be
\label{Hcm}
H_{\rm cm} (\tilde{\omega})= T_{\rm cm} + {1\over 2 }mA\tilde{\omega}^2R_{\rm cm}^2
-{3\over 2}\hbar\tilde{\omega} \ .  
\ee 

Here, $\tilde{\omega}$ is a free parameter and not necessarily
identical to the frequency $\omega$ of the underlying oscillator
basis. The generalized center-of-mass Hamiltonian~(\ref{Hcm}) exhibits
a zero-energy Gaussian ground-state wave function for all values of
$\tilde{\omega}$. In what follows, we demonstrate that the
coupled-cluster ground state is the zero-energy eigenstate of $H_{\rm
cm}(\tilde{\omega})$ for a suitably chosen frequency
$\tilde{\omega}$. Thus, the coupled-cluster wave
function factorizes, and the center-of-mass wave function is a
Gaussian. 

We denote the expectation value of the generalized center-of-mass
Hamiltonian~(\ref{Hcm}) in the coupled-cluster ground state as
\be
\label{Ecm}
E_{\rm cm}(\tilde{\omega}) \equiv \langle H_{\rm cm}(\tilde{\omega})\rangle \ .
\ee 
We compute the expectation value~(\ref{Ecm}) via the Hellmann-Feynman
theorem, i.e. we add a small perturbation $\beta H_{\rm cm}(\omega)$
with $\beta=0.001$ to the intrinsic Hamiltonian~(\ref{Hin}), and the
expectation value results from the difference 
quotient~\cite{Dierksen,Enzerhof}. 

First, we consider the standard center-of-mass Hamiltonian and set
$\tilde{\omega}=\omega$. The results are shown in the inset of
Fig.~\ref{fig1}. Comparing the results for the center-of-mass energy
$E_{\rm cm}(\omega)$ with the ground-state energy suggests that the
latter does not depend on the former. This is a first hint that the
intrinsic and center-of-mass coordinates decouple. For the model space
with $\hbar\omega\approx 20$~MeV the expectation value $E_{\rm
cm}(\omega)$ vanishes approximately, thus indicating that the wave
function factorizes. As a check, we fix $\hbar\omega=20$~MeV, add the
term $\beta H_{\rm cm}(\omega)$ to the intrinsic Hamiltonian, and
compute the ground-state energy of the resulting Hamiltonian. We find
that the ground-state energy varies by merely 15~keV as $\beta$ is
increased from zero to one. Since the intrinsic energy shown in
Fig.~\ref{fig1} is practically independent of the frequency of the
underlying oscillator basis, the wave function must approximately
factorize for all values of $\hbar\omega$ shown in
Fig.~\ref{fig1}. However, only for $\hbar\omega\approx 20$~MeV, do we
know that the center-of-mass wave function is a Gaussian.

\begin{figure}[h]
\includegraphics[width=0.37\textwidth,clip=]{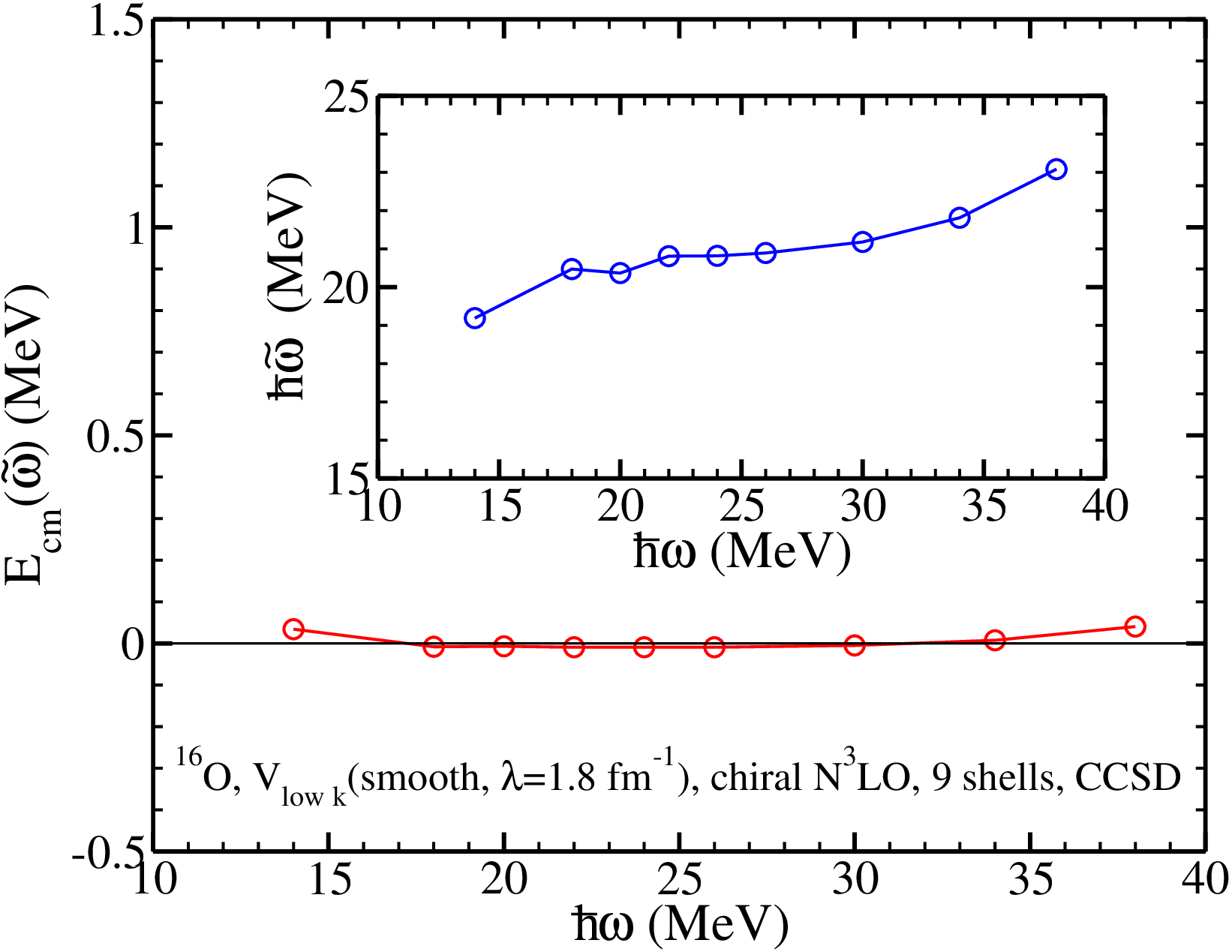}
\caption{(Color online) $^{16}$O ground-state expectation value $E_{\rm
cm}(\tilde{\omega})$ (within CCSD) of the generalized center-of-mass
Hamiltonian $H_{\rm cm}(\tilde{\omega})$ as a function of the
oscillator spacing $\hbar\omega$. The model space consists of nine
major oscillator shells. Inset: Relation between the frequency
$\tilde{\omega}$ and the frequency $\omega$ of the underlying
oscillator basis. }
\label{fig2}
\end{figure}

Let us assume that the center-of-mass wave function generally has a
Gaussian shape, i.e. it is the ground state of the generalized
center-of-mass Hamiltonian~(\ref{Hcm}) for a suitably chosen frequency
$\tilde{\omega}$. It is thus our task to determine this frequency.
To this purpose, we employ the identity
\ba
H_{\rm cm}(\omega)+{3\over 2}\hbar\omega-T_{\rm cm} = 
{\omega^2\over\tilde{\omega}^2}\left(H_{\rm cm}(\tilde{\omega})
+{3\over 2}\hbar\tilde{\omega}-T_{\rm cm}\right) \ ,\nonumber
\ea
take its expectation value, require $E_{\rm cm}(\tilde{\omega})=0$, employ $\langle T_{\rm cm}\rangle={3\over 4}\hbar\tilde{\omega}$, and determine the unknown
frequency $\tilde{\omega}$ from the already computed expectation values $E_{\rm
cm}(\omega)$. This yields the two possible frequencies
\ba
\label{magic}
\hbar\tilde{\omega} = \hbar\omega + {2\over 3} E_{\rm cm}(\omega) \pm
\sqrt{{4\over 9}(E_{\rm cm}(\omega))^2 +{4\over 3}\hbar\omega E_{\rm
cm}(\omega)} \ .  
\ea 
We employ these frequencies in the generalized center-of-mass
Hamiltonian and compute the corresponding expectation values $E_{\rm
cm}(\tilde{\omega})$. We find that one expectation value is close to
zero, while the other is usually very large.  The small expectation
values are shown in Fig.~\ref{fig2}, and the corresponding frequencies
$\tilde{\omega}$ are shown in the inset of Fig.~\ref{fig2}. The
expectation values are very small compared to the energy
$\hbar\tilde{\omega}\approx 20$~MeV of spurious center-of-mass
excitations.  The practically vanishing expectation values demonstrate
that the coupled-cluster wave function factorizes.  This is the main
result of this Letter.  The reader should not be concerned about the
fact that some of the expectation values shown in Fig.~\ref{fig2}
assume small negative values (of size -0.01~MeV). Recall that the
coupled-cluster method is non-variational when the cluster operator is
truncated, and that small negative expectation values are thus
tolerable.  As shown in the inset of Fig.~\ref{fig2}, the frequency
$\tilde{\omega}$ corresponding to the Gaussian center-of-mass wave
function varies only little as the frequency $\omega$ of the
underlying oscillator basis is changed. We made the following two
checks. First, we computed the expectation value of $T_{\rm cm}$ and
found that $T_{\rm cm}\approx {3\over 4}\hbar\tilde{\omega}$, as
expected for a Gaussian. Second, we repeated the calculations directly
in the oscillator basis and did not employ the Hartree-Fock basis. We
again find very small values for $E_{\rm cm}(\tilde{\omega})$ of order
-0.1~MeV, and an almost constant frequency $\tilde{\omega}$ very
close to what we found in the Hartree-Fock basis.

Let us also consider interactions with higher momentum cutoff. We
employ Entem and Machleidt's chiral nucleon-nucleon
interaction~\cite{N3LO} at next-to-next-to-next-to-leading order
(N$^3$LO). This interaction has an approximate high-momentum cutoff of
$\lambda\approx 500$~MeV.  We apply the coupled-cluster method to
compute the ground-state energy of $^{16}$O. Due to the relatively
high momentum cutoff of the interaction, the wave function is more
correlated. This requires us to employ a large model space and
three-particle-three-hole cluster amplitudes to obtain converged
solutions.  The model spaces consist of $N=19$ oscillator shells;
however, the maximal single-particle orbital angular momentum is kept
at $l\le 13$. We have verified that this is sufficient for a
convergence of the results. For the three-particle-three-hole
clusters, we employ the $\Lambda$-CCSD(T)
approximation~\cite{Kuch,Taube}.

\begin{figure}[h]
\includegraphics[width=0.37\textwidth,clip=]{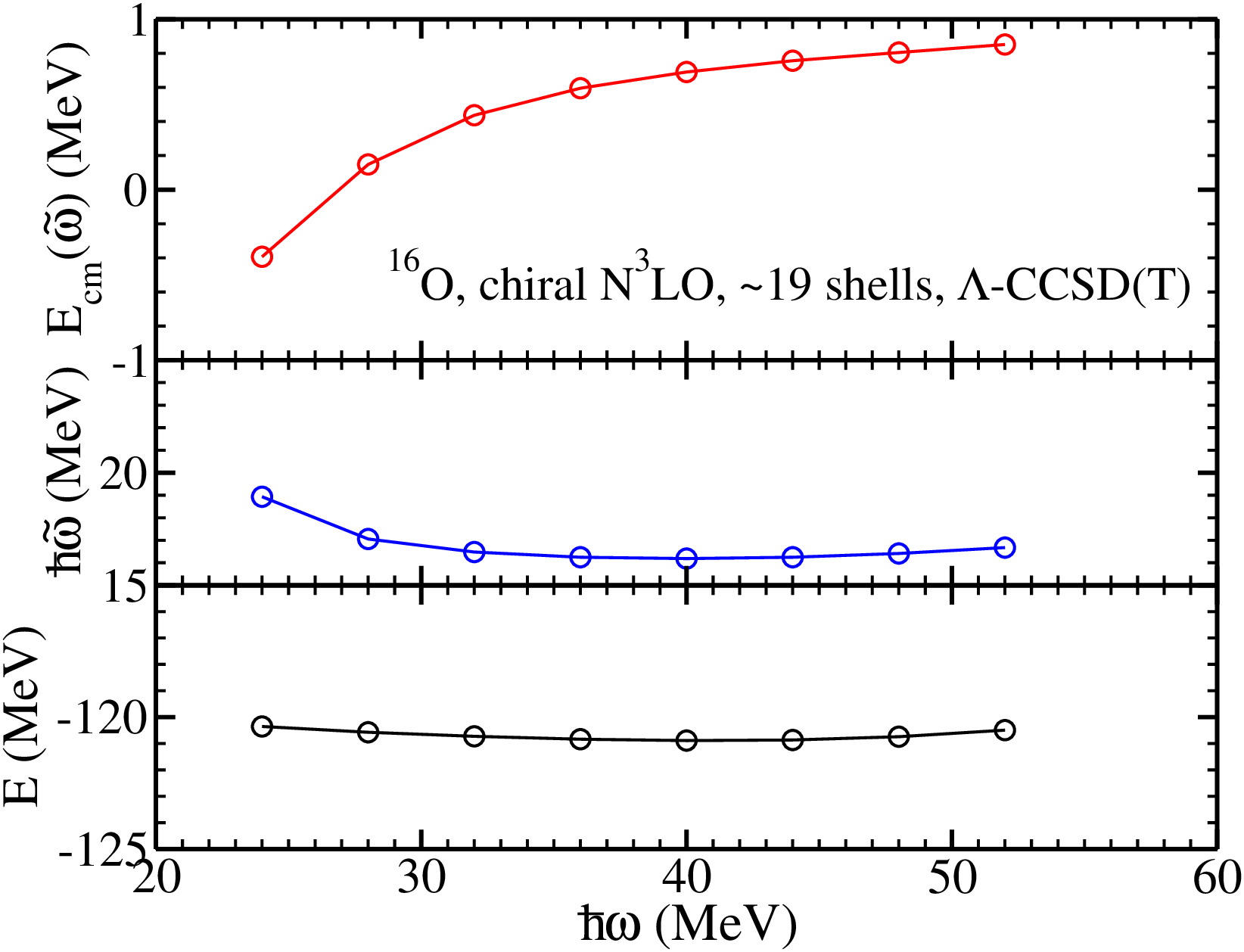}
\caption{(Color online) Bottom: ground-state energy of $^{16}$O within
the $\Lambda$-CCSD(T) approximation as a function of the frequency
$\hbar\omega$ of the underlying oscillator basis. Middle: Relation
between the frequency $\tilde{\omega}$ and the frequency $\omega$ of
the underlying oscillator basis. Top: Expectation value $E_{\rm
cm}(\tilde{\omega})$ of the center-of-mass vs. $\hbar\omega$.}
\label{fig3}
\end{figure}

The bottom part of Fig.~\ref{fig3} shows that the ground-state energy
is practically independent of the oscillator frequency of the
underlying single-particle basis for a large frequency range. This
demonstrates that the results are well converged with respect to the
size of the model space. We find that the expectation value $E_{\rm
cm}(\omega)$ of the standard center-of-mass Hamiltonian increases with
increasing frequency $\omega$ of the model space, and that it assumes
values of tens of MeV and varies strongly with the frequency $\omega$
of the underlying oscillator basis. Clearly, the intrinsic
ground-state energy is independent of the expectation value $E_{\rm
cm}(\omega)$, and this suggests again a decoupling of the intrinsic and
the center-of-mass wave functions.

Let us again assume a Gaussian shape for the center-of-mass wave
function and follow our two-step procedure. First, we employ
Eq.~(\ref{magic}) and compute the two possible frequencies
$\tilde{\omega}$ that are consistent with the already computed
expectation value $E_{\rm cm}(\omega)$. Second, we compute the two
corresponding expectation values $E_{\rm cm}(\tilde{\omega})$. As
before, we find a large and a small expectation value for each
frequency $\omega$ of the underlying oscillator basis. The small
expectation values are shown in the top part of Fig.~\ref{fig3}. The
corresponding frequencies $\tilde{\omega}$ are shown in the middle
part of Fig.~\ref{fig3}. While not as impressive as for the
low-momentum interaction, the expectation values $E_{\rm
cm}(\tilde{\omega})$ are below 1~MeV in size, and much smaller than the
binding energy of $^{16}$O or any of its excitations. In particular,
the expectation values are small compared to the energy
$\hbar\tilde{\omega}\approx 16$~MeV of the spurious
center-of-mass excitations. A simple two-level model yields that the
wave function has about 6\% squared overlap with spurious
states. The relative small negative eigenvalues are again due to the
non-variational character of the coupled-cluster method, and they are
tolerable within the overall accuracy of the calculation. Thus, we can
conclude that the coupled-cluster wave function also exhibits an
approximate factorization~(\ref{factor}) for the interaction with a
relatively high momentum cutoff.  We speculate that quadruple cluster
amplitudes would be necessary to further reduce the magnitude of
$E_{\rm cm}(\tilde{\omega})$. Note that the frequency $\tilde{\omega}$
that determines the width of the Gaussian center-of-mass wave function
is again almost constant over a wide range of frequencies of the
underlying oscillator basis.

\begin{figure}[h]
\includegraphics[width=0.37\textwidth,clip=]{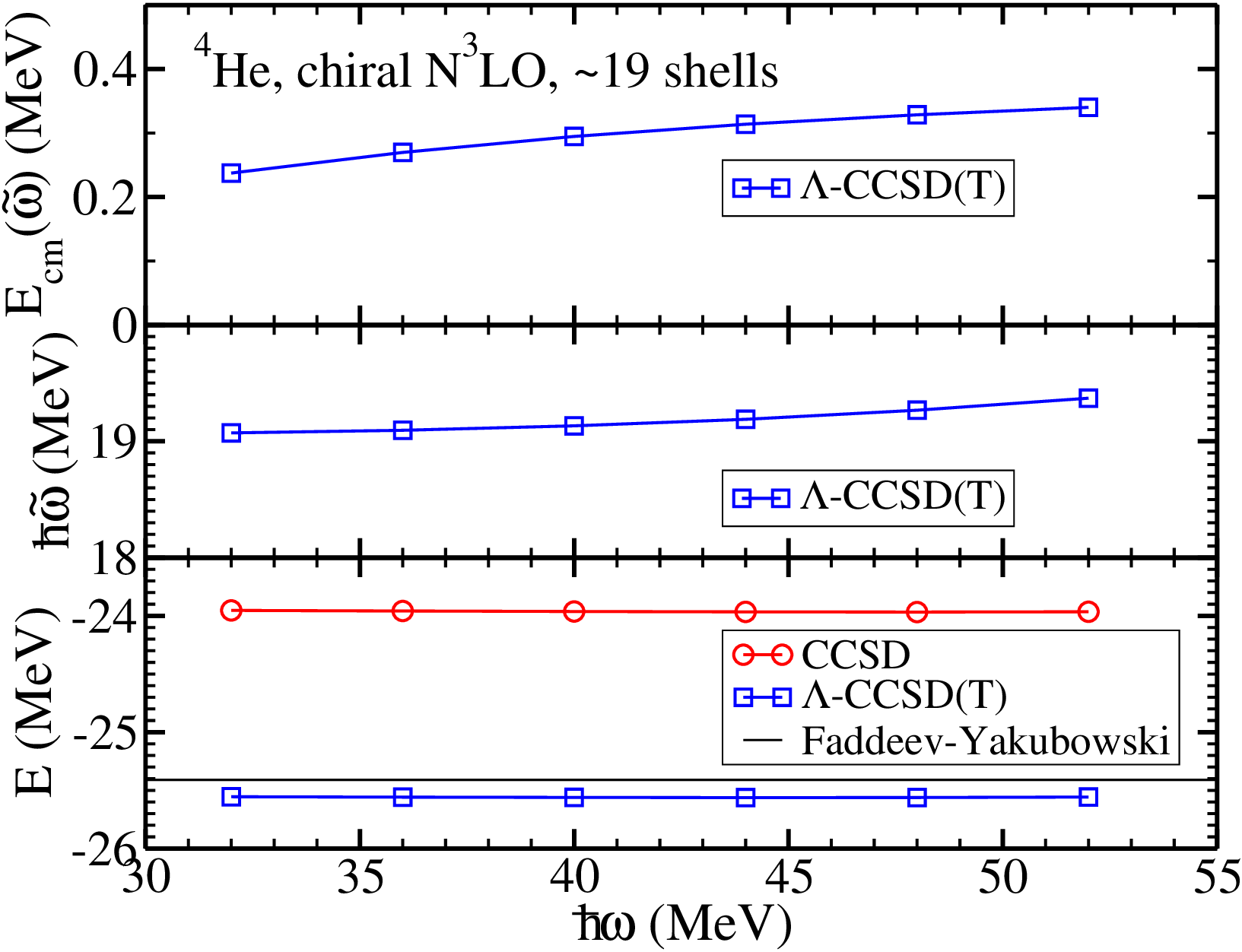}
\caption{(Color online) Same as Fig.~\ref{fig3}, but for $^4$He.}
\label{fig4}
\end{figure}

Finally, we employed the chiral potential also for $^4$He and show the
results in Fig.~\ref{fig4}. The bottom part shows the ground-state
energy ($E\approx-25.56$~MeV in $\Lambda$-CCSD(T)) and compares it to
virtually exact Faddeev-Yakubowski calculations ($E\approx -25.41$~MeV \cite{NavCau}).
The middle part shows the frequency of the
approximate Gaussian center-of-mass wave function, while the top part
shows the expectation value $E_{\rm cm}(\tilde{\omega})$. Again, the
approximate factorization is very statisfactory. At this moment, we
have no profound understanding of the observed factorization.

In summary, we presented strong numerical evidence that wave function
of the atomic nucleus is a product of an intrinsic and a center-of-mass
wave function in a sufficiently large model space {\it even} when the
model space is not a complete $N\hbar\omega$ oscillator space. The
center-of-mass wave function is approximately a Gaussian whose width
varies little with the frequency $\omega$ of the underlying oscillator
basis. The reported results open the door for a verifiable description
of translationally invariant states for a large variety of model
spaces and many-body methods.

We acknowledge discussions with P. Navr{\'a}til, R. Roth, and
S. Quaglioni that motivated this research. We thank the Institute for
Nuclear Theory at the University of Washington for its hospitality
during the program {\it Effective Field Theories and the Many-Body
Problem}. This work was supported by the U.S. Department of Energy
under Contract Nos. \ DE-AC05-00OR22725 with UT-Battelle, LLC (Oak
Ridge National Laboratory (ORNL)), under Grant No.\ DE-FG02-96ER40963
(University of Tennessee (UT)), and under DE-FC02-07ER41457 (UNEDF SciDAC
Collaboration). This research used computational resources of the
National Institute for Computational Sciences (UT/ORNL) and the
National Center for Computational Sciences (ORNL).

\end{document}